\documentclass[aps, prl,reprint, twocolumn] {revtex4-1}

\usepackage{subcaption}
\usepackage{csquotes}
\usepackage{graphicx}

\usepackage{mathtools}
\usepackage{amsmath}
\usepackage{siunitx}
\usepackage{dsfont}
\usepackage{amssymb}
\usepackage{amsfonts} 
\usepackage{graphicx}
\usepackage{dcolumn}
\usepackage{bm}
\usepackage{textcomp}
\usepackage{color}
\usepackage[normalem]{ulem}

\newcommand*\mean[1]{\bar{#1}}

\newcommand{\thi}{\mean{\tau}_\mathrm{high}}
\newcommand{\tlo}{\mean{\tau}_\mathrm{low}}
\newcommand{\drts}{\Delta_\mathrm{RTS}}
\newcommand{\wn}{\sigma_\mathrm{wn}}
\newcommand{\wnratio}{Q_\mathrm{wn}}
\newcommand{\wnest}{\sigma_{\mathrm{wn}_\mathrm{est.}}}
\newcommand{\pn}{\sigma_\mathrm{pn}}
\newcommand{\pnratio}{Q_\mathrm{pn}}

\newcommand{\ntraps}{N_\mathrm{traps}}
\newcommand{\npeaks}{N_\mathrm{peaks}}
\usepackage{ragged2e}
\DeclareCaptionJustification{justified}{\justifying}
\begin{document}

\title{Extensive Study of Multiple Deep Neural Networks for \\
 Complex Random Telegraph Signals}

\author{Marcel Robitaille$^{1,2,3}$, HeeBong Yang$^{1,2,3}$, Lu Wang$^{1}$, Na Young  Kim$^{1,2,3,4,5,6*}$}
\address{$^1$Department of Electrical and Computer Engineering, University of Waterloo,  200 University Ave W,  Waterloo,  N2L 3G1,  Ontario, Canada}
\address{$^2$Institute for Quantum Computing,  University of Waterloo,  200 University Ave W,  Waterloo,  N2L 3G1,  Ontario, Canada}
\address{$^3$Waterloo Institute for Nanotechnology, University of Waterloo,  200 University Ave W,  Waterloo,  N2L 3G1,  Ontario, Canada}
\address{$^4$Department of Physics and Astronomy, University of Waterloo,  200 University Ave W,  Waterloo,  N2L 3G1,  Ontario, Canada}
\address{$^5$Department of Chemistry, University of Waterloo   200 University Ave W,  Waterloo,  N2L 3G1,  Ontario, Canada}
\address{$^6$Perimeter Institute, 31 Caroline Street N,  Waterloo, N2L 2Y5, Ontario, Canada}

\email{nayoung.kim@uwaterloo.ca}

\date{\today}

\begin{abstract}
Time-fluctuating signals are ubiquitous and diverse  
in many physical, chemical, and biological systems, among which random telegraph signals (RTSs) refer to a series of instantaneous switching events between two discrete levels from single-particle movements. Reliable RTS analyses are crucial prerequisite to identify underlying mechanisms related to performance sensitivity. When numerous levels partake, complex patterns of multilevel RTSs occur, making their quantitative analysis exponentially difficult, hereby systematic approaches are found elusive. Here, we present a three-step analysis protocol via progressive knowledge-transfer, where the outputs of early step are passed onto a subsequent step. Especially, to quantify complex RTSs, we build three deep neural network architectures that can process temporal data well and demonstrate the model accuracy extensively with a large dataset of different RTS types affected by controlling background noise size. Our protocol offers structured schemes to quantify complex RTSs from which meaningful interpretation and inference can ensue.
\end{abstract}

\maketitle

\begin{figure*} 
	\includegraphics[width=0.9\linewidth]{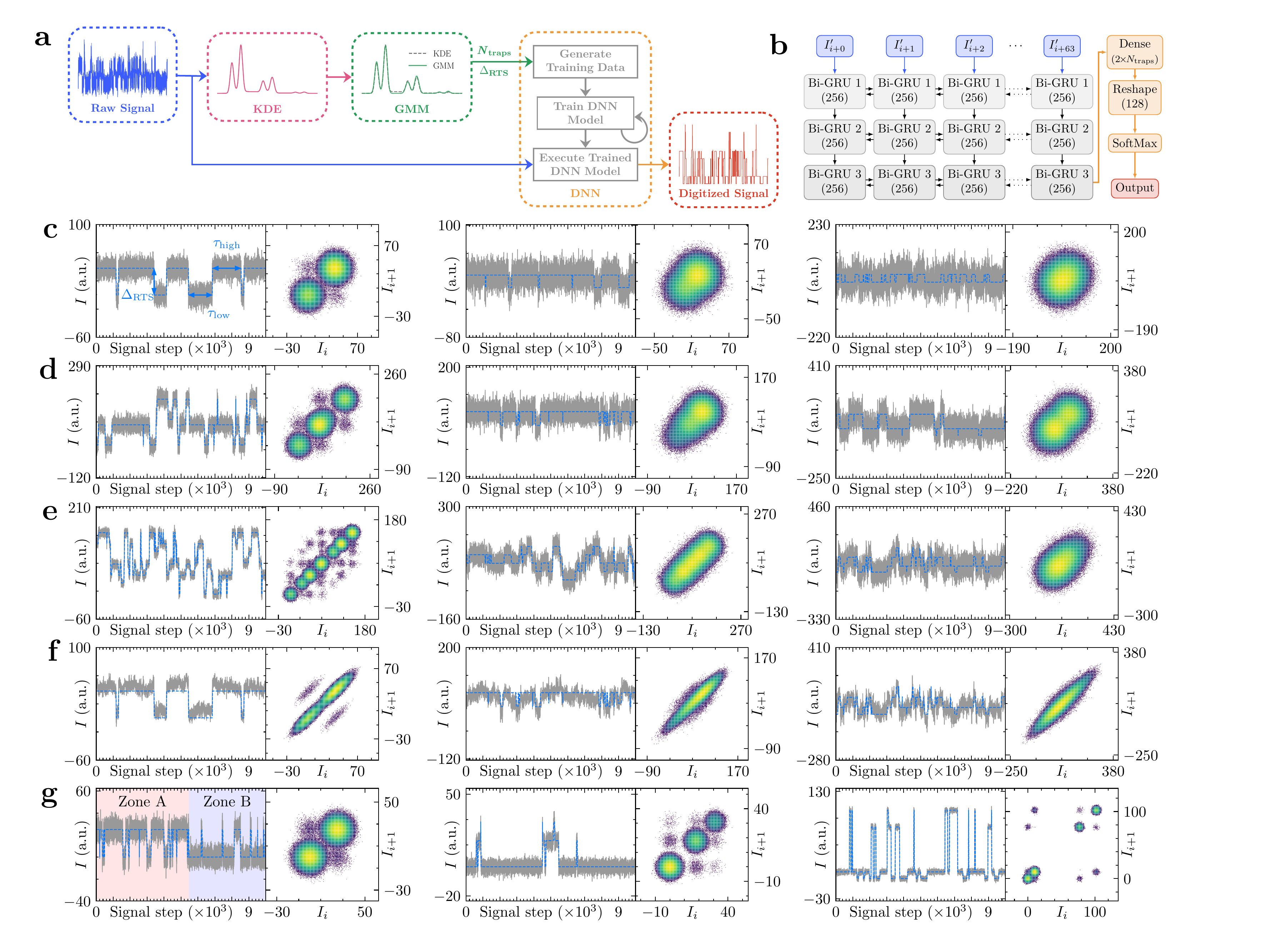}
	\caption{\small \textbf{Types of simulated random telegraph signals and a flow chart of a three-step RTS analysis protocol.}
	\textbf{a,} A flowchart representing the flow of data through our algorithm with visualizations for various steps.
	\textbf{b,} The structure of our recurrent neural network, where Bi-GRU is the standard bidirectional gated recurrent unit from keras.
	\textbf{c-e,} The time-series and time-lag plots for exemplary signals from normal random telegraph signals (nRTSs) of \textbf{(c)} 1-trap, \textbf{(d)} 2-trap, and  \textbf{(e)} 3-trap with Gaussian white noise levels $\wnratio=\SI{20}{\percent}$ (left),  $\wnratio=\SI{60}{\percent}$ (middle), and $\wnratio=\SI{100}{\percent}$ (right). The definition of $\wnratio$ is given in Methods. \textbf{f,} Representative nRTSs with $1/f$ (pink) noise in place of white noise for 1-trap with $\wnratio=$\SI{20}{\percent} pink noise (left), 2-trap and \SI{60}{\percent} pink noise (middle), 3-trap and \SI{100}{\percent} pink noise (right). \textbf{g,} The time-series and time-lag plots for exemplary signals with three anomalous RTS types: a metastable signal with the two distinctive zones (A and B) annotated and color-coded (left), a missing-level signal where it is possible to see only 3 clusters on the diagonal of the time-lag plot (middle), and a coupling signal where it is possible to see that the amplitude of the fast trap changes with the state of the slow trap (right). The blue digitized signals are obtained from the Bi-GRU model in all time-domain signals.}
	\label{fig:fig1}
\end{figure*}

Random telegraph signals (RTSs) are sequential data that comprise a time series of burst signals between two well-defined states. As a fundamental type of low-frequency stochastic fluctuations~\cite{Hooge_1981,Dutta_1981,Kleinpenning_1990,Howard_2017}, RTSs appear in a wide variety of disciplines: physics, chemistry, biology, and engineering. Two-level RTSs are manifestation of a single-carrier stochastic movement that can switch abruptly between high and low levels separated by $\drts$ with individual characteristic dwell times $\tau_{\text{high}}$ and $\tau_{\text{low}}$ defined in a representative signal (Fig.~\ref{fig:fig1}\textbf{c}), whose mean values of $\thi$ and $\tlo$ obey Poisson statistics~\cite{Theodorsen_2017}. These distinctive RTSs are frequently seen in small-sized electronic devices such as metal-oxide-semiconductor field-effect transistors~\cite{Lundberg_noisesources,Kogan1996,Hung1990, Yang2022}, resistive random access memory devices~\cite{Zahoor2020}, memristors~\cite{Hu2021,Li2021}, and CMOS image sensors~\cite{Wang2006}. For example, in a submicron-sized MOSFET, if there is a trap to capture or emit a single charge carrier, device currents in the transport channel exhibit arbitrary patterns of two-level RTSs over time. Similar RTS phenomena are discerned in superconducting qubits~\cite{Lambert2020}, single photon avalanche diodes~\cite{Capua2021}, ultrasensitive biosensors~\cite{Ribeiro2020}, and single-cell activities in ion channels~\cite{Liebovitch1987}. Real-time dynamics in quantum electrical devices also reveal rapid jumping sequences~\cite{Kenfack2017, Kleinherbers2022}. Furthermore these bursting events can represent biological
and biochemical processes like spontaneous transitions in gene-regulation network~\cite{Kepler2001,Tsimring2014}, where RTS models are adopted to explain stochastic individual gene activation and deactivation processes well. Therefore, thorough analysis of RTSs is essential to quantify performance sensitivity of devices and to deepen fundamental understanding of systems and processes in diverse areas. 

Tremendous efforts have been put to establish RTS analysis methods that obtain quantitative values of three parameters $\drts$, $\thi$, and $\tlo$ for each RTS. A basic way to determine $\drts$ is from a histogram of all data points; a clear bimodal distribution in the histogram confirm the two levels of RTS, consequently giving its $\drts$ value. $\thi$ and $\tlo$ are computed as the statistical mean of each time duration if RTSs display as digitized data. A barrier to estimate RTS parameters accurately is other unwanted noise such as white Gaussian noise or pink noise, which mask the RTSs. An efficient solution to this barrier is a time-lag plot (TLP) that depicts not only level-corresponding peaks but also between-level transitions~\cite{Nagumo_2009}, and a weighted TLP further improves peak-detection efficiency by minimizing unwanted broad noise~\cite{Martin-Martinez2014}. Notwithstanding the mixture of RTS and other background noise, a simple hidden Markov model (HMM) has been widely used to analyze two-level RTSs by assigning a pair of two-by-two matrices to represent hidden layers associated with transition probabilities between the states. HMM performs remarkably well in obtaining $\drts$ and digitizing the signals into the binary levels~\mbox{\cite{Rabiner1989, Puglisi2012, Li2013}}.

As the physical size of devices further shrinks down to nanoscales or movements of a single particle play important roles, complex RTS patterns among multiple discrete levels of several traps have often been captured~\cite{Awano_2013, Nagumo_2010}. Figure~\ref{fig:fig1}\textbf{c-g} showcase representative noisy signals with two-level and multi-level RTSs. As the number of RTS-levels increases, the total number of parameters to determine each RTS surges exponentially, where the simple HMM fails to estimate all parameters successfully because digitizing noisy signals into multiple discrete levels is extremely challenging. Hence, more advanced factorial HMMs (fHMM) with many hidden layers were proposed and attempted to evaluate multi-level RTSs, exploiting its capacity to handle complex sequential data~\mbox{\cite{puglisi_random_2013,puglisi_rtn_2013,francesco_maria_puglisi_factorial_2014}}.

Although fHMMs assess complex RTSs well, to set up a fHMM is not straightforward but daunting because it is not clear how the initial values of the matrices defining the hidden layers of the fHMM should be decided. Worst, the number of all matrix elements rises steeply as more discrete levels are involved. Hence, systematic analysis models of complex RTSs are sought-after, and to the best of our knowledge, a step-by-step protocol for the quantitative RTS analysis is lacking. Furthermore, its validation through extensive statistical analysis of RTS parameters with a large dataset has been elusive except the work by \citeauthor{Awano_2013} who evaluated full parameters using Bayesian inference only for eight multi-trap RTS examples within limited statistical analysis of accuracy, where structured study of additional background noise is missing~\cite{Awano_2013}. Here we demonstrate a three-step RTS analysis protocol based on various machine learning (ML) algorithms and deep learning architectures (See Methods). Our protocol is validated thoroughly with a large synthetic dataset that include various types of RTS signals on top of background noise, and it is used to analyze real current signals from carbon nanotube devices.

\section{Results}

The first two steps determine two quantities of $\ntraps$ and $\drts$ that become initial values to generate large training data together with featureless background noise for mimicking raw signals in practice. These generated data are utilized to train deep neural network (DNN) models built upon three advanced architectures that are specifically designed to process sequential data: (1) bi-directional gated recurrent unit (Bi-GRU)-based recurrent neural networks (RNNs), (2) bi-directional long short-term memory (Bi-LSTM)-based RNNs, and (3) WaveNet convolutional neural networks~\cite{oord_2016}. As the flowchart indicates in Fig.~\ref{fig:fig1}\textbf{a}, we emphasize that the trained DNN models execute directly the original raw signals (blue in Fig.~\ref{fig:fig1}\textbf{a}) to produce digitized signals (red in Fig.~\ref{fig:fig1}\textbf{a}), wherein our novelty lies.

In order to acquire the statistical validation of our RTS analysis protocol and to compare the execution performance among three different DNN architectures, we generate a multitude of simulated signals, each having 1,000,000 time steps in total. The number of traps ($\ntraps$) in the simulated signals selects 1, 2, and 3. Consequently, the associated number of discrete RTS levels are 2, 4, and 8, respectively. Moreover, we consider normal and anomalous RTSs. The former result from the superposition of multiple ($\ntraps$) independent two-level RTSs (Fig.~\ref{fig:fig1}\textbf{c-f}), whereas the latter exhibit complex interaction schemes between traps, namely, the traps are not independent in the anomalous case (Fig.~\ref{fig:fig1}\textbf{g}). In addition to variations on simulated the RTS itself, we add two different types of background noise: Gaussian white noise and pink noise, each masking the underlying RTS and impeding the analysis. In the following, we present the results from the algorithm performance validation study on synthetic normal RTS (nRTS) and anomalous RTS (aRTS) with multiple background noise sizes. Furthermore, we examine real RTSs from short-channel carbon nanotube film devices.

The performance results are summarized quantitatively in terms of the RTS amplitude ($\drts$) and digitization ($\eta$) whose error is calculated by,

\begin{align}
	\epsilon(X) &\equiv \frac{|{X}_\mathrm{theory} - {X}_\mathrm{estimate}|}{{X}_\mathrm{theory}} \times \SI{100}{\percent},
\end{align}
where ${X}_\mathrm{theory}$ is the design value for parameters ($X\in\{\drts,\eta\}$) of the simulated RTS and ${X}_\mathrm{estimate}$ is the output from our algorithm analysis. Because the time constants $\tlo$ and $\thi$ can be extracted from the digitized signals, the error trends of $\tlo$ and $\thi$ are correlated with $\epsilon(\eta)$, which is hence presented below.

\begin{figure} 
	\includegraphics[width=\linewidth]{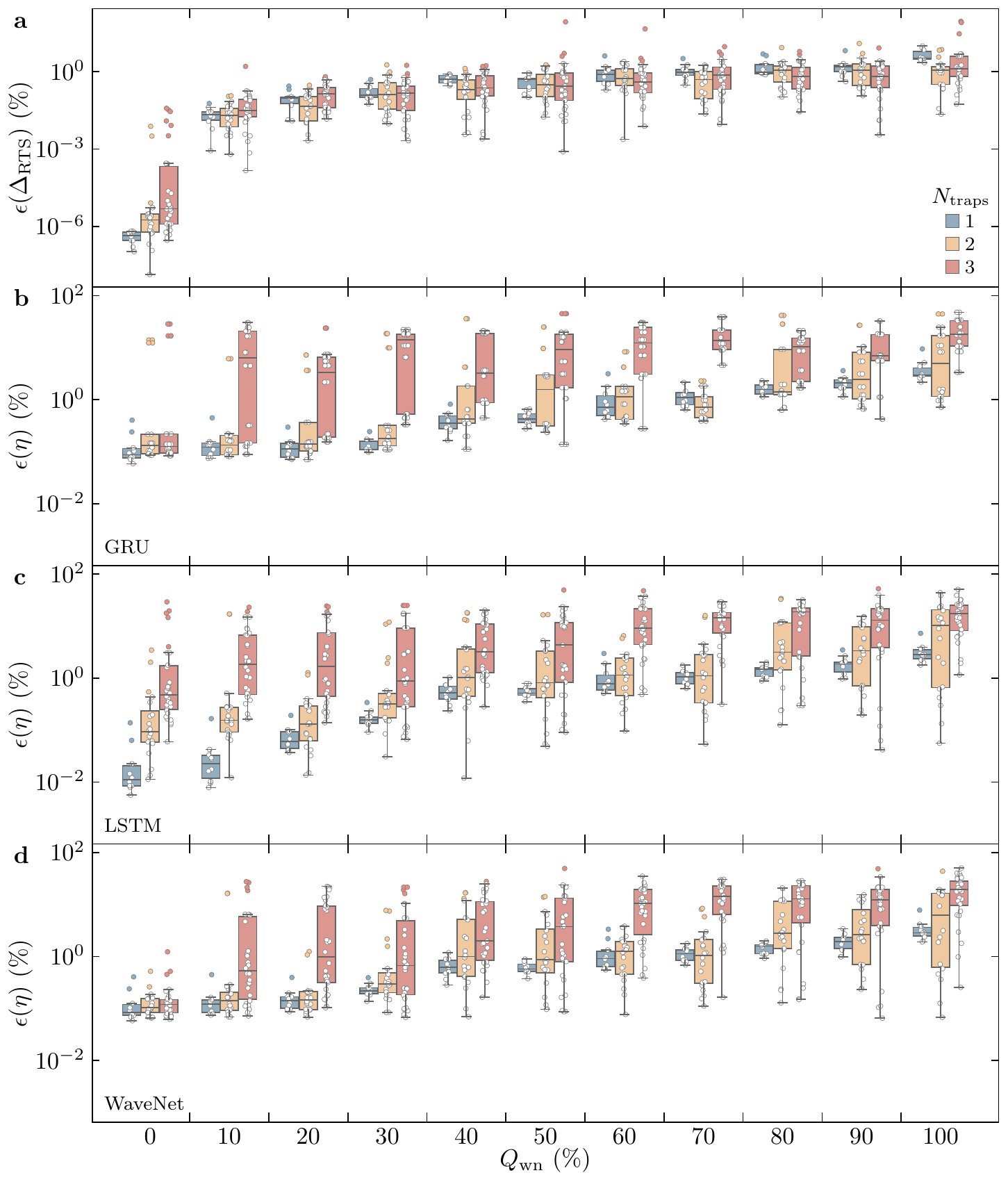}
	\caption{\small\textbf{Validation results of nRTSs with white noise background.} \textbf{a,} $\epsilon(\drts)$, the error in predicting $\drts$ from the KDE and GMM steps are plotted in the logarithmic scale. The digitization error $\epsilon(\eta)$ are computed by three trained deep neural network models: \textbf{b,} bi-directional gate recurrent units (GRU);\textbf{c,} bi-directional long short-term model (LSTM);\textbf{d,} WaveNet for $\ntraps=$ 1 (blue), 2 (yellow), and 3 (red) examples. Note that each point corresponds individual RTS, so in each $\wnratio$ bin, the boxplot of $\ntraps=$ 1 contains 10 RTS pattern examples, but those of $\ntraps=$ 2 and 3 are from 20 and 30 RTS patterns.}%
	\label{fig:fig2nRTSwn}
\end{figure}

\begin{figure} 
	\includegraphics[width=\linewidth]{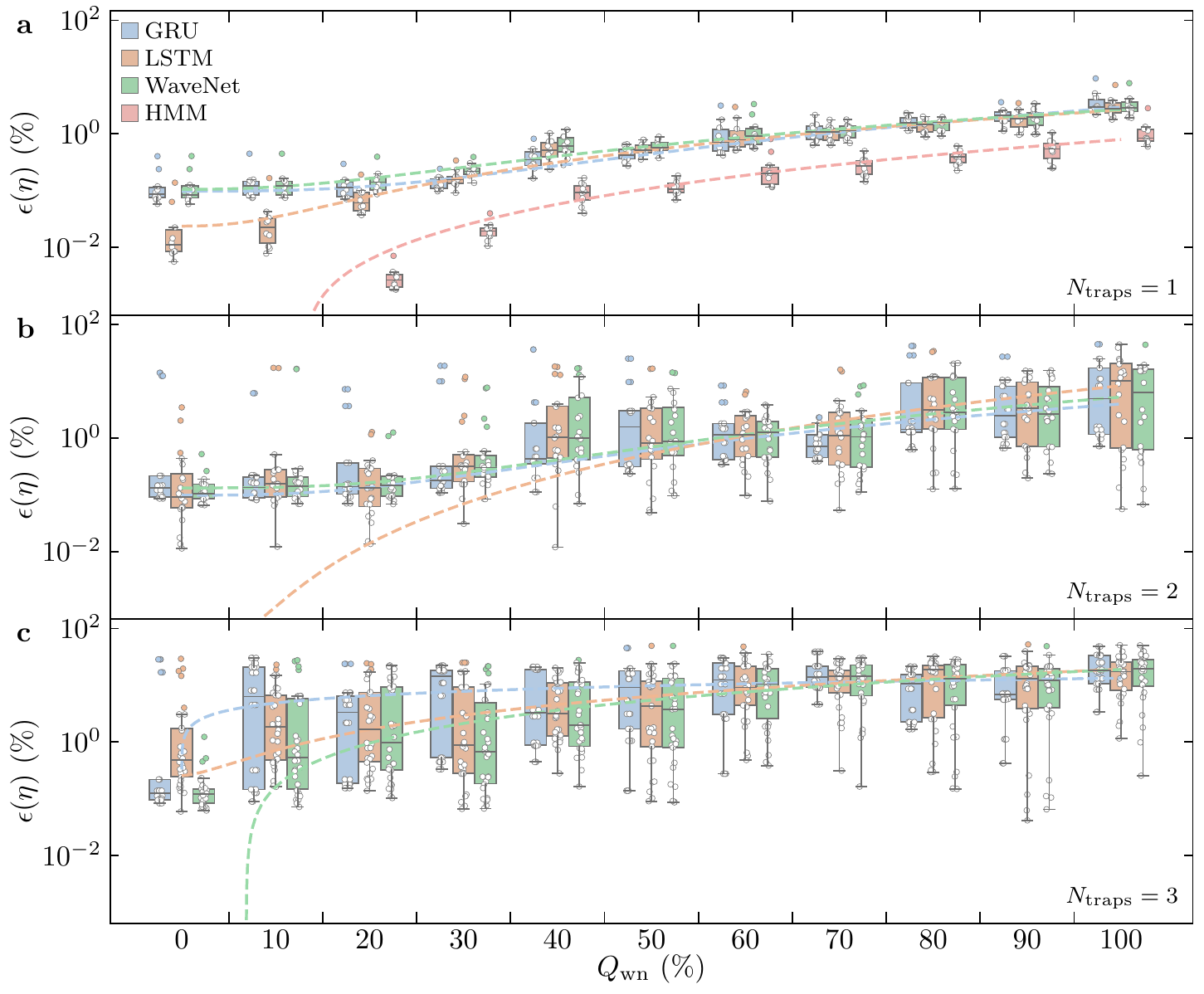}
	\caption{\small\textbf{Comparison of digitization error boxplots from deep neural network models. \textbf{a,}} $\epsilon(\eta)$ as a function of $\wnratio$ for all 110 examples from our nRTS white noise dataset where $\ntraps$ = 1 for models Bi-GRU (blue), Bi-LSTM (orange), WaveNet (green), as well as HMM (red). $\epsilon(\eta)$ where $\ntraps$ = 2 (\textbf{b}) and 3 (\textbf{c}) for three DNN models with the same color scheme.}%
	\label{fig:fig3dnn}
\end{figure}

\subsection*{Validation on synthetic normal random telegraph signals with white noise}

First, we apply our three-step analysis protocol to all 330 nRTSs synthesized for $\ntraps \in \{1,2,3\}$ that are veiled by 11 sizes of the background white noise amplitudes, $\wnratio \in \{0\,\%,10\,\%,20\,\%,\dots, 100\,\%\}$ (See Methods). Figure~\ref{fig:fig2nRTSwn} collects all error analysis results over the entire dataset. The first KDE step followed by the second GMM estimate $\ntraps$ and $\drts$, with which the amplitude error $\epsilon(\drts)$ is calculated. Our algorithm detects $\ntraps$ correctly for all 220 examples (100\,\% success) where $\ntraps\in\{1,2\}$, but only for 107 of the 110 (97.3\,\% success) 3-trap RTS examples. The three failed examples of 3-trap RTS occur when the amplitudes of two traps are very close and the difference of these amplitudes is easily obscured by background noise, which can hamper the recognition of individual peaks. Two failures occur in the incidence of the worst class, $\wnratio = 100\,\%$.

We choose boxplot graphs to present explicitly the distribution of $\epsilon(\drts)$ along $\wnratio$, giving median and interquartile range for $\ntraps$ --- 1 (blue), 2 (yellow), and 3 (red) --- side by side in Fig.~\ref{fig:fig2nRTSwn}\textbf{a}. As $\ntraps$ is elevated and $\wnratio$ escalates, it is increasingly difficult to estimate accurate $\drts$ values, resulting in higher $\epsilon(\drts)$. Indeed, the difficulty of this task increases exponentially with $\ntraps$ due to the increased number of KDE peaks (RTS levels) by $\npeaks = 2^{\ntraps}$. A more important factor to $\epsilon(\drts)$ is the influence of background noise expressed by $\wnratio$, which is are explicitly correlated to the spread of the peak distributions clearly seen in the TLPs (Fig.~\ref{fig:fig1}\textbf{c-g}). In contrast to the examples where $\wnratio = 20\,\%$ (left column), the background noise completely hides individual peaks in the examples where $\wnratio = 100\,\%$ (right column), producing a rather broad distribution in the TLPs that lowers the peak resolving accuracy. Surprisingly, $\epsilon(\drts)$ remain below $\SI{2}{\percent}$ for all three traps when $\wnratio \leq \SI{40}{\percent}$ and it is impressive that even for the worst class of $\ntraps = 3$ and $\wnratio = \SI{100}{\percent}$, the third quartile sits below \SI{10}{\percent} error. This indicates that our KDE and GMM steps can deliver accurate estimates of $\ntraps$ and $\drts$ with confidence, which play a crucial role in the subsequent step where these values are used to generate tailored training data for our DNN model.

The third step aims to digitize RTSs that are degraded by background noise and to recover the pure RTS patterns. This process consists of several progressive tasks: differentiating RTS patterns from background noise, recording well-defined discrete states belonging to RTSs, setting decision criteria to assign each data point to one of the discrete states, and executing all time-series data in sequential or parallel manners. In validation, we quantify the digitization error $\epsilon(\eta)$ of the digitized RTSs at the third step with respect to the ground truth time series. Figure~\ref{fig:fig2nRTSwn}\textbf{b-d} plot $\epsilon(\eta)$ for all nRTSs as a function of $\wnratio$ with the trained Bi-GRU, Bi-LSTM, and WaveNet DNN models, respectively. Several common trends are captured in the $\epsilon(\eta)$ versus $\wnratio$ boxplots of all three DNN models. First of all, $\epsilon(\eta)$ is much higher than $\epsilon(\drts)$ in all examples, reflecting the difficulties in digitization. The behavior of $\epsilon(\eta)$ versus $\ntraps$ and $\wnratio$ is expected: higher $\epsilon(\eta)$ for increased $\ntraps$ and $\wnratio$ is similar to the $\epsilon(\drts)$ trend. However, the degradation in $\epsilon(\eta)$ is much more dramatic. When $\ntraps$ is 1, $\epsilon(\eta)$ grows rather semilogarithmically along $\wnratio$, whereas for $\ntraps$ = 2 and 3, $\epsilon(\eta)$ are high in all ranges of $\wnratio$ in the Bi-GRU and the WaveNet DNNs.  

In order to compare the three DNN models directly, we plot their $\epsilon(\eta)$ side by side given $\ntraps$ in Fig.~\ref{fig:fig3dnn}.  For $\ntraps$ = 1, we also include the HMM result after correcting the aforementioned flip error as a reference. In all $\wnratio$, Bi-GRU and WaveNet DNNs exhibit similar statistical tendency; however, the Bi-LSTM works better for $\wnratio$~= 0 and 10\,\% about 10 times lower $\epsilon(\eta)$ than those of two counterparts. All three DNNs show comparable performance for the difficult cases of $\ntraps$ = 1 and all  of $\ntraps$ = 2 and 3. The three models have the third quartiles of $\epsilon(\eta)$ only below 10\,\% error up to $\wnratio$ = 90\,\% for $\ntraps$ = 2, which is an encouraging result. 

We inquire whether there is a common mathematical asymptotic trend between $\epsilon(\eta)$ and $\wnratio$. We attempt to perform a power-law fitting with the form of $\wnratio^{\beta}$  where $\beta$ is the power exponent. In Fig.~\ref{fig:fig3dnn}, the fitting results are in the dashed line from the median of each bin. For $\ntraps$ = 1 and 2, the $\beta$ values are between 2.3 and 3.9, which means that our algorithm yields small $\epsilon(\eta)$ for less than 40\,\% $\wnratio$ but suffers with large $\wnratio$. The $\ntraps$ = 3 examples are apparently different in the entire $\wnratio$ ranges so that the $\beta$ is relatively lower between 0.4 and 1.8. 

\subsection*{Validation on synthetic normal random telegraph signals with pink noise}

In addition to white noise, we examine the effect of another common background fluctuation: pink noise. This is a form of low-frequency noise because its power spectral density (PSD) follows a trend of $1/f^\alpha$, where $f$ is the frequency and $\alpha$ is the power-law exponent. We replace previous Gaussian white noise with pink noise whose PSD follows the curve $1/f^{1}$. The pink noise magnitude in relation to the RTS magnitude is denoted $\pnratio$, which has a similar definition to $\wnratio$. For $\pnratio$ below 40\,\%, the influence of white noise and pink noise look alike, for example, in the cases of $\wnratio = $ 20\,\% in Fig.~\ref{fig:fig1}\textbf{c} and $\pnratio = $ 20\,\% in the left panel of Fig.~\ref{fig:fig1}\textbf{f}. However, higher $\pnratio$ RTSs (Fig.~\ref{fig:fig1}\textbf{f}, middle and right) display not only rapid wiggles but also slow-varying, incidental background trends that bury the RTSs immensely in the time domain. For the 330 RTS examples with pink noise, the same algorithm struggles both to accurately extract $\ntraps$ and $\drts$ in the KDE and GMM steps and to recover the pure RTS patterns in the digitization step.

\begin{figure} 
	\includegraphics[width=\linewidth]{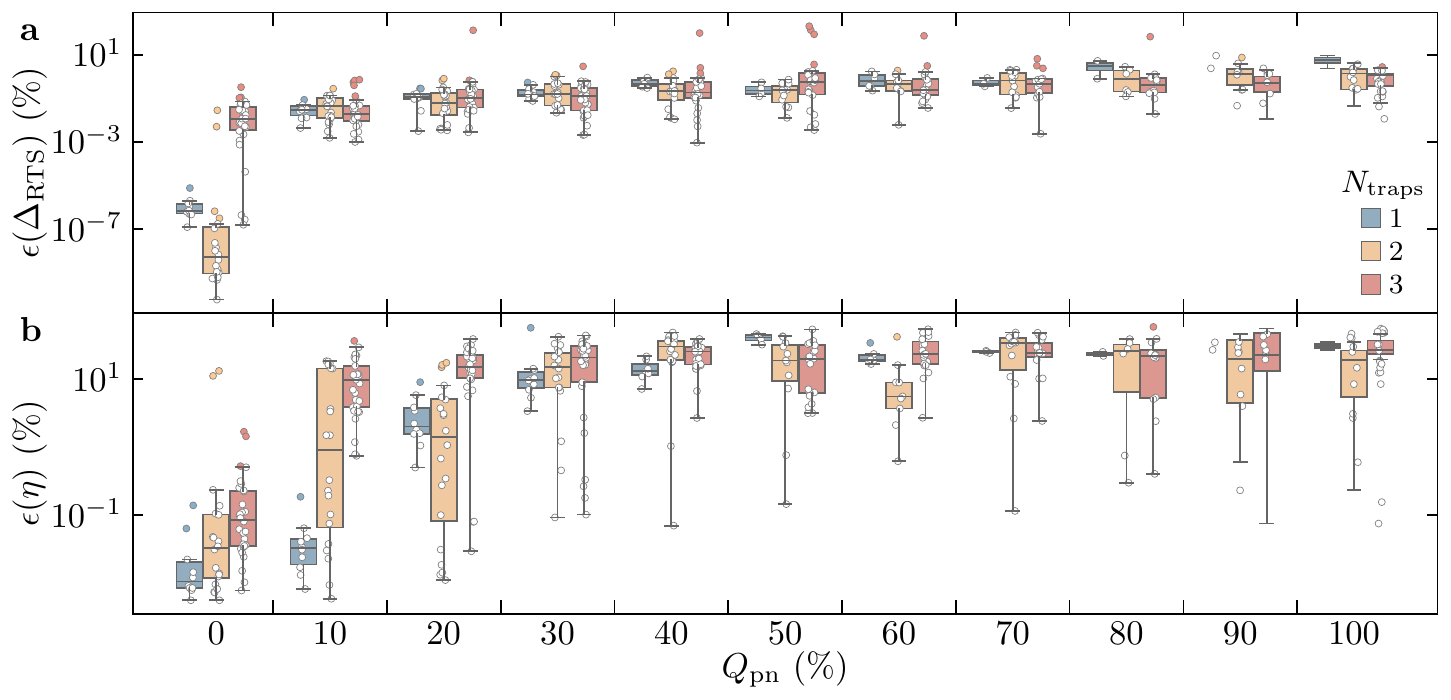}
	\caption{\small\textbf{Validation results of the Bi-GRU model on normal RTS with pink noise.} The error in predicting $\drts$ (\textbf{a}) and the digitized signal (\textbf{b}) against $Q_\text{pn}$ for $\ntraps= 1$ (blue), 2 (yellow), and 3 (red) over our normal RTS dataset with pink noise.}%
	\label{fig:fig4nRTSpn}
\end{figure}

To analyze the pink noise effect on all 330 nRTS examples, we focus on the Bi-GRU RNN architecture which is fully connected and a newer RNN than the Bi-LSTM. When $Q_\text{pn}$ is less than 30\,\%, the correct number of traps is detected for all examples and $\epsilon(\drts)$ is on the same order as that of the white noise examples in the previous section. This means that our algorithm can recognize multilevel RTSs well for up to 30\,\% of both white and pink noise. On the other hand, if $Q_\text{pn} \geq 40\,\%$, we failed to accurately resolve all traps and their RTSs. Since our method uses $\ntraps$ and $\drts$ from prior steps to generate training data for our DNN models, $\epsilon(\eta)$ is only quantified for detected traps. When $Q_\text{pn} \geq 40\,\%$, some or all traps are missed for many examples, which are omitted from Fig.~\ref{fig:fig4nRTSpn}. As long as the KDE and GMM steps find $\ntraps$ and $\drts$, the accuracies of both amplitude and digitization are as good as those of the white noise case, which is one important lesson. However, the mean of $\epsilon(\eta)$ is always higher than that of the white noise case. We anticipate that some filtering techniques prior to our current algorithm may help to lower both $\epsilon(\drts)$ and $\epsilon(\eta)$ which should be thoroughly studied in further studies. 

\subsection*{Validation on synthetic anomalous random telegraph signals}

\begin{figure} 
	\includegraphics[width=\linewidth]{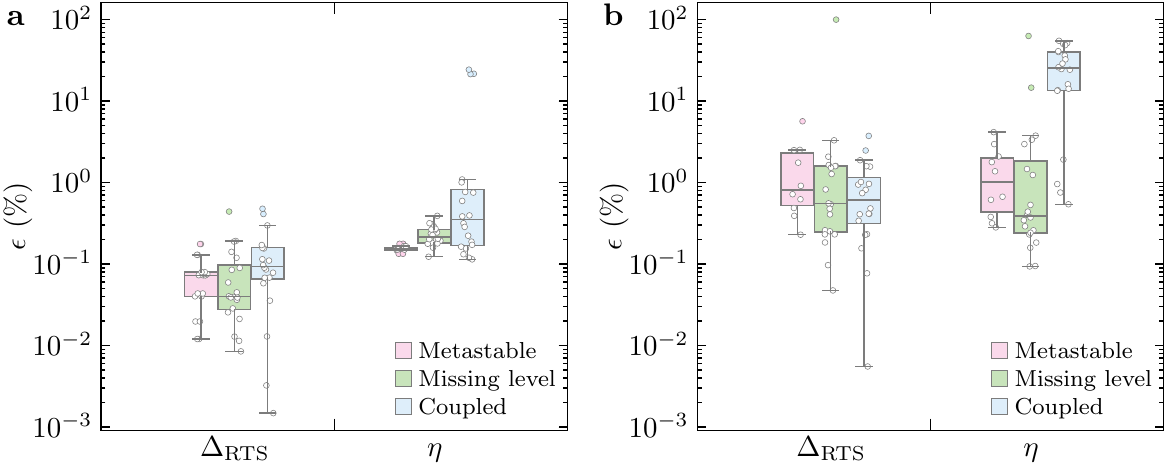}
	\caption{\small\textbf{Validation results of the Bi-GRU model on anomalous RTS.}  \textbf{a,} $\epsilon(\drts)$ and $\epsilon(\eta)$ for three aRTS types degraded by white noise with $\wnratio$ = 20\,\%. \textbf{b,} $\epsilon(\drts)$ and $\epsilon(\eta)$ where white noise is replaced by pink noise with $\pnratio = 20\,\%$.}%
	\label{fig:fig5aRTS}
\end{figure}

We expand the validation of our algorithm further by testing it on aRTSs and the results are presented in Fig.~\ref{fig:fig5aRTS}. These signals may look somewhat similar to nRTSs, but there are some noticeable characteristics in time-series signals, often making aRTSs harder to address. Varieties of these aRTSs are grouped into three categories with distinct features~\cite{zhang_comprehensive_2018,Wang2018}. The first group is labelled as \textquote{metastable} aRTS, where a reversal two-level RTS for $\ntraps=1$ forms two different zones A and B in the left of Fig.~\ref{fig:fig1}\textbf{f}. The origin is associated with a metastable state, which stays in the low state dominantly ($\tlo > \thi$) during zone A, while the high state has longer dwell time ($\thi > \tlo$) in zone B given the same trap. The other two categories are associated with multi-trap aRTSs, and we take $\ntraps=2$ for demonstration. When two traps are strongly coupled, the switching of one trap depends exclusively on the state of other. For example, one trap is only active when the other is also turned on. In this case, there are only three discrete levels instead of four, which is an example of \textquote{missing-level} aRTS (Fig.~\ref{fig:fig1}\textbf{f}, middle). The third type shows full four-level RTSs, but the switching amplitude $\drts$ of each trap depends on the state of the other (Fig.~\ref{fig:fig1}\textbf{f}, right), whose category is named \textquote{coupled} aRTS.

For simplicity, we analyze 10 simulated signals of each aRTS type with fixed $\wnratio =~{\SI{20}{\percent}}$ as well as the same aRTS with $\pnratio ={\SI{20}{\percent}}$ pink noise with the bi-GRU RNN.
Figure~\ref{fig:fig5aRTS}\textbf{a} displays the results of this aRTS white noise study, where the mean amplitude extraction error is only {\SI{0.1}{\percent}}, which is comparable with the result from the nRTSs and $Q_\text{wn}~=~20\,\%$. Although $\epsilon({\eta})$ is around 10 times higher than that of the nRTS counterpart, it is below 1\,\% except for three outliers in the weak coupling case. The pink noise again makes the aRTS problems harder, increasing error by 10 times for both $\drts$ and digitization,
which is presented in Fig.~\ref{fig:fig5aRTS}\textbf{b}. Given this promising result from a limited dataset, we aspire to investigate aRTS thoroughly, furthermore, we consider introducing a designated step for each aRTS type that can facilitate to find an optimized algorithm in future work. 

\subsection*{Random telegraph signals of carbon nanotube devices }
\begin{figure} 
	\includegraphics[width=\linewidth]{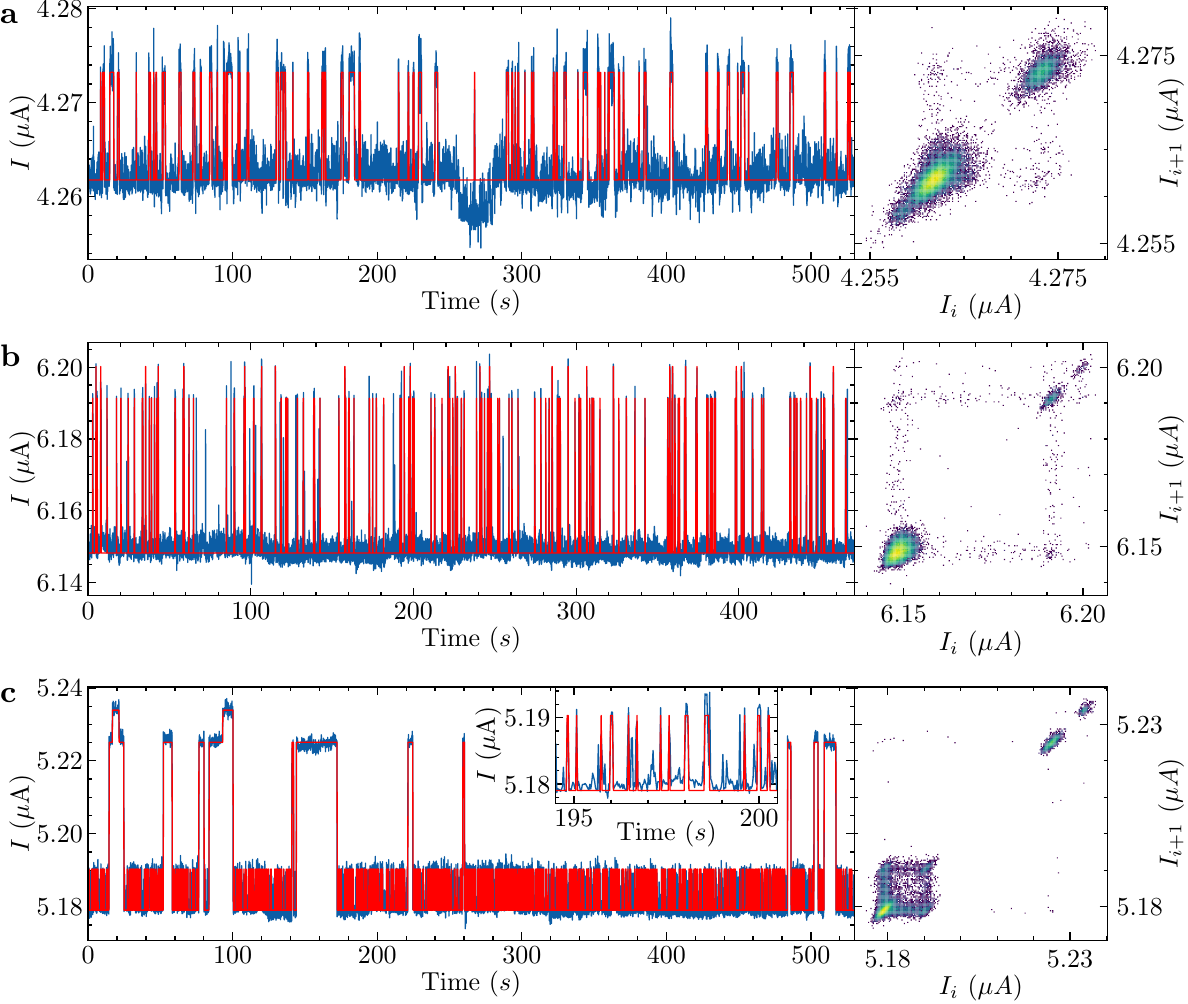}
	\caption{\small\textbf{Analysis results of the Bi-GRU on real RTS data.} In all three panels, the digitization result (red) are overlaid onto the raw measurement of a carbon nanotube film (blue) in the left panel and the time-lag plot of the raw signal are in the right panel. \textbf{a,} We assign this signal as a two-level nRTS with low-frequency background noise. \textbf{b,} An example of a missing level aRTSs is given. \textbf{c,} Three-trap RTSs with different dwell time scales are seen as a complicated aRTSs. The inset shows a zoomed-in view of the fast trap.}%
	\label{fig:fig6cnt}
\end{figure}

The extensive validation tests on numerous synthetic RTSs grant confidence in our algorithm. However, this study would not be complete without examining real experimental data. The device under test has two terminals and a {\SI{65}{\nano\meter}} thick, {\SI{500}{\nano\meter}} wide multi-walled carbon nanotube film acting as a channel. The current through this channel is measured at {\SI{9}{\kelvin}}. In in Fig.~\ref{fig:fig6cnt}, three representative raw signals with increasing complexity are selected in blue  and overlaid in red are the digitized signals successfully obtained from our algorithm. Figure~\ref{fig:fig6cnt}\textbf{a} and \textbf{b} are a simple two-level nRTS example  and two mutually exclusive traps as a missing-level aRTS category, respectively. More complicated data are examined in Fig.~\ref{fig:fig6cnt}\textbf{c}  where aRTS-like features are observed. One small amplitude trap is only active when the large amplitude trap is inactive; however, the other small amplitude trap is only active when the large-amplitude trap is also active. First, our analysis teaches that at least three traps are involved and some interactions among them exist. When we treat the data in Fig.~\ref{fig:fig6cnt}\textbf{c} in the nRTS category, the resulting digitization was poor. Instead, we find that a missing-level aRTS analysis provides an extremely well-fitting digitization result. Unlike the synthetic RTSs, the accuracy concept is not valid in real data because the ground truth values are not known, but our algorithm provides quantitative information of RTS parameters as a prerequisite for researchers to draw appropriate interpretation about their own systems.

\section*{Discussions}

The quantitative analysis of complex RTSs is essential to understand devices, systems, or processes in many scientific and engineering fields because they can reveal principal mechanisms and limit sensitivity. HMMs are extremely efficient to digitize two-level RTSs regardless of strong background noise, but their big issue is a flip error, namely, the labels of discrete two states are completely swapped, technically, 100\,\% error for around 50\,\% of examples. While this flip error can be easily corrected in the simple case, it seems not to be straightforward for complex RTSs among many levels and level-couplings. In addition, the extension of HMMs or fHMMs into multi-level RTSs is not clear due to intrinsic uncertainty in sophisticated HMM, which becomes a big hurdle to assess complex multi-level RTSs widely. 

Our analysis protocol is structured in three steps and one of its key contributions is the progressive knowledge transfer approach, exploiting the knowledge discovered in early steps to generate tailored training data in the final step. A major advantage in our method over HMMs is that we predict the switching events of each trap independently, the superposition of which being the digitized multi-level RTS, whereas the latter predicts a single multi-level RTS. In the KDE and GMM steps, primary error sources are undesirable noise on top of RTSs. Therefore, $\epsilon(\drts)$ grows against $\wnratio$ and $\pnratio$. Strangely, we notice the non-negligible $\epsilon(\drts)$ in nRTS even for $\wnratio=0$\,\% and $\pnratio=0$\,\% which seem counter-intuitive. We attribute the $\drts$ estimation error to the resolution of the KDE. Indeed, while the KDE is theoretically continuously, the implementation used operates finite input and output spaces, as is often the case for the computer implementations of mathematical concepts. The length and range of this space determine the finite resolution of the KDE in practice. 
 
Numerous sources of error worsen digitization accuracy, such as incorrectly choosing the RTS level between two close values, missing a switching event, or including a false switching event. The occurrence of these faults becomes more frequent in intricate, long-sequence RTSs with high $\ntraps$ and background noise. Moreover, it is found that the information of neighboring data at prior and subsequent time steps can facilitate the decision of the state assignment. 
Recognizing RTSs as a sequence labeling task with the same input and output length, we approach ubiquitous RTS problems by resorting to powerful DNN architectures specifically designed for time-sequential data. Bi-GRU and Bi-LSTM RNNs are built and trained, where forward and backward propagation can include correlations of nearby data points to improve the analysis. We also examine another powerful network architecture known as Wavenet, a form of convolutional neural network (CNN)~\cite{oord_2016}. 

We summarize the training complexity of three DNN models in Table 1 (see Methods) and examine the training times of the three DNN models with a Titan Xp graphics card, 504\,GB of memory, and 28 CPU cores. For $\ntraps = 1$, we find that the simple HMM can solve each example within a duration of on average 1.8\,s for $\wnratio$ = 0\,\%, but around 20\,s for  $\wnratio$ = 100 \%, linearly increasing against $\wnratio$ values. Meanwhile, all DNN models take much longer than the HMM computation duration for $\ntraps$ = 1. Thus, the HMM is a clear choice to examine two-level RTSs with the condition of the flip-error correction. However, we get an important observation that all three DNN models have reasonably constant training time regardless of $\ntraps$, $\wnratio$, $\pnratio$, or the normal vs. anomalous cases. The training times of the Bi-GRU, bi-LSTM, and WaveNet models are 1350 $\pm$ 1 s, 1570 $\pm$ 1s, and 897 $\pm$ 1 s. In terms of both the trainable parameters and training time, the CNN WaveNet model would be simpler and faster in the RTS analysis given our synthetic data. This should be further investigated whether this come from the architecture difference between convolution layers and recurrent schemes. The insensitivity of the training time to parameters supports that the DNN models possess potential advantages and capacity to handle much more complicated RTSs beyond our synthetic data. 

In conclusion, we establish a structured methodology to analyze complex RTSs step by step with three DNN architectures, whose performance is directly validated with various types of RTSs and the effect of other noise to collect the accuracy statistics of RTS parameters for the first time to our knowledge. We assert that our DNN-based algorithm based is well-suited to characterize multi-level RTS in diverse fields,
such as the study of nano- or micro-scale solid-state devices, quantum devices at cryogenic temperatures, chemical processes and biological systems.

\section*{Methods}

\subsection*{Systematic three-step algorithm}
The algorithm is comprised of three major steps: first, the probability density of the raw signal is calculated and $\ntraps$ is estimated; second, a GMM is fitted to the previous probability density to extract $\drts$ of each trap; and third, these values of $\ntraps$ and $\drts$ are used to generated tailored training data for the DNN model to reconstruct the pure RTS.

\textbf{\textit{Step 1: Peak Estimation.}}
The goal of this step is to recognize $\ntraps$ present in the raw, noisy signal and to estimate the location of each peak in the probability density, which is a seed value for the GMM\@. To uncover peaks masked by white noise~\cite{Smith_1999}, we first apply a moving average. Although the moving window width $M$ is a tuning parameter, we successfully scale $M$ by an estimate of the noise in the signal with the expression $M=\mathrm{floor}[60 \cdot \tan^{-1}(\sigma_{\mathrm{wn}_\mathrm{est.}} / 4) / \pi]$, which gives excellent results over our dataset.

Then, we create a density distribution of the filtered data using KDE with a Gaussian kernel. Contrary to a histogram, the KDE is a continuously differentiable kernel function that leads to a smoother density distribution. Our Gaussian KDE reduces the number of false peaks caused by randomly fluctuating white noise, but maintains the ability to resolve close peaks.

\textbf{\textit{Step 2: Amplitude Extraction.}}
In the presence of Gaussian noise, we can represent the probability density function (PDF) of an isolated RTS component by the mixture model of two Gaussian functions: $G_\mathrm{high} = q \cdot G(\mu=I_0 + \drts, \sigma)$ and $G_\mathrm{low} = p \cdot G(\mu=I_0,\sigma)$ for high and low levels of the RTS\@, where $p = \frac{\tlo}{\thi + \tlo}$, $q = (1-p)$, and $G$ is a standard Gaussian function with parameters mean ($\mu$) and standard deviation ($\sigma$). The total probability distribution $G_T$ for multi-trap RTS is constructed via the convolution of individual $G_\mathrm{high}$ and $G_\mathrm{low}$ pairs~\cite{Parzen_1960}. Finally, non-linear least-squares curve-fitting is applied to fit this model to the PDF obtained from the KDE\@. The fitted parameters of $G_T$ return the $\drts$ of each trap.

\textbf{\textit{Step 3: Pure RTS Reconstruction.}}

In this final step, the problem of reconstructing the masked RTSs is translated to a multi-output classification problem, which we solve with RNNs since these are well-suited for time-sequence data~\mbox{\cite{lecun2015deep}}. The input to the model is the time-sequence of a noisy RTS normalized between 0 and 1 and the outputs are time-series predictions for the state of each trap from the set $\mathbb{S} = \{\text{high},\text{low}\}$.

The RNN model is based on a 3-layer bidirectional gated recurrent unit (Bi-GRU) network~\mbox{\cite{cho2014learning,deng2019sequence}} with 256 neurons per layer (Fig.~\ref{fig:fig1}\textbf{b}). The input Bi-GRU layer takes historical information allowing a comprehensive analysis. With a window size of $W=64$, time steps 0--63 predict the RTS level at step 63, then the window is shifted to predict step 64 using steps 1--64, and so on. The last Bi-GRU layer yields a high-dimensional tensor including historical data. To obtain a single predicted RTS level per time step, this high-dimensional tensor is fed through a dense layer with softmax activation, whose output is reshaped to 3 dimensions sized $(W, \ntraps, {\mathbb{S}}=2)$. Finally, the argmax function selects the most confidently predicted state $\in\mathbb{S}$ at each time step, and the desired time-series signal for each trap is acquired~\mbox{\cite{goodfellow2016deep}}.

RTS analysis is difficult to solve with supervised machine learning as there are no perfectly labelled measurement data with which to train. We solve this by training an RNN model for each example under test on twenty \num{50000}-point synthetic RTS signals tailored for the specific $\ntraps$, $\drts$, and $\wnest$ which are quantified in previous steps. These data are shuffled and divided using a {$\SI{80}{\percent}/\SI{20}{\percent}$} train/validation split.

In addition to GRU-based RNNs, we also test LSTM and WaveNet. In the case of LSTM, only the recurrent layers changed and the network architecture is otherwise identical. The WaveNet structure is created according to \citeauthor{oord_2016}~\cite{oord_2016}. The number of trainable parameters and training time for each of these is given in Table~\ref{tab:trainable_parameters} as an indicator for the architecture complexity.

\begin{table}[h]
	\caption{Training complexity of GRU, LSTM, and WaveNet architectures}%
	\label{tab:trainable_parameters}
	\centering
	\begin{ruledtabular}
	\begin{tabular}{ccc}
		\textbf{Architecture} & \textbf{\# Trainable Parameters} & \textbf{Training time} \\
		\hline
		GRU & $\num{695018} + \ntraps \cdot \num{65792}$ & $\SI{1350}{\second}\pm\SI{1}{\second}$ \\
		LSTM & $\num{922114} + \ntraps \cdot \num{65792}$ & $\SI{1570}{\second}\pm\SI{1}{\second}$ \\
		WaveNet & $\num{40130} + \ntraps \cdot \num{8256}$ & $\SI{897}{\second}\pm\SI{1}{\second}$ \\
	\end{tabular}
	\end{ruledtabular}
\end{table}

The servers used for training the machine learning models are equipped with Titan Xp or Titan V graphics cards, 128-504\, GB of memory, and 24-28 CPU cores.

\subsection*{Data generation for validation test}

For validating our algorithm with statistical analysis, we synthesize noisy signals of 1,000,000 time-steps by summing two contributions: digitized RTSs and other background noise. For nRTSs, three  $\ntraps$ values of 1,2, and 3 are considered, whereas for aRTSs, depending on the classifications, $\ntraps$ is 1 or 2.  The individual RTS of each trap is generated by iteratively drawing high and low dwell times from a geometric distribution with the time-constant parameters $\thi$ or $\tlo$. 

As featureless fluctuating background noise, white noise is introduced and its amplitude distribution obeys the Gaussian probability density function,
\begin{align}
	f_\text{G}(I) = \frac{1}{\wn \sqrt{2\pi}} \exp\left(-\frac{I^2}{2\wn^2}\right),
\end{align}
where $\wn$ is the standard deviation of the white noise. Since both the white-noise amplitude and their $\wn$ degrade RTSs, $\wnratio$ is defined to be the ratio between $\wn$ and the smallest switching amplitude ($\min(\Delta_{\text{RTS},i})$) of any trap per example as follows: $$\wnratio \equiv \frac{\wn} {\min\limits_{i\in\{1, \dots, \ntraps\}}(\Delta_{\text{RTS}, i})},$$ where $i$ is a trap index. In this way, $\wnratio$ is effectively the inverse of the signal-to-noise ratio, taking one of 11 values from $\SI{0}{\percent}, \SI{10}{\percent}, \SI{20}{\percent}, \ldots, \SI{100}{\percent}$. Another prevalent background noise source considered is $1/f$ noise often called pink noise, which obeys Gaussian distributions but has a power law spectrum with a power exponent 1. The pink noise amplitude sizes are parameterized by, 
$$Q_\text{pn} \equiv \frac{\pn} {\min\limits_{i\in\{1, \dots, \ntraps\}}(\Delta_{\text{RTS}, i})},$$
where $i$ is a trap index.

In practice, white Gaussian noise and pink Gaussian noise is generated using the Python \texttt{numpy} package~\cite{numpy} and the Python \texttt{colorednoise} package~\cite{colorednoise}, respectively. Given a configuration of ($\ntraps, \wnratio$), 10 different incidences are prepared for the dataset of the total 330 nRTSs in the presence of white noise or pink noise and the total 30 aRTSs in the presence of white noise or pink noise.

\section*{Acknowledgements}
We acknowledge the support of Industry Canada, the Ontario Ministry of Research \& Innovation through Early Researcher Awards (RE09-068), and the Canada First Research Excellence Fund-Transformative Quantum Technologies (CFREF-TQT). 

\bibliography{RTSDNNrefs}

\end{document}